\documentstyle[twocolumn,aps,floats,amssymb,epsfig]{revtex}

\begin{document}

\draft
\tolerance = 10000

\renewcommand{\topfraction}{0.9}
\renewcommand{\textfraction}{0.1}
\renewcommand{\floatpagefraction}{0.9}
\setlength{\tabcolsep}{4pt}

\twocolumn[\hsize\textwidth\columnwidth\hsize\csname @twocolumnfalse\endcsname

\title{Phase Diagram for the Winfree Model of Coupled Nonlinear
Oscillators}
\author{Joel T. Ariaratnam and Steven H. Strogatz}
\address{Center for Applied Mathematics, Cornell University, Ithaca,
New York 14853}
\date{08 December 2000}

\maketitle

\begin{abstract}
In 1967 Winfree proposed a mean-field model for the spontaneous
synchronization of chorusing crickets, flashing fireflies,
circadian pacemaker cells, or other large populations of
biological oscillators. Here we give the first bifurcation
analysis of the model, for a tractable special case.  The system
displays rich collective dynamics as a function of the coupling
strength and the spread of natural frequencies. Besides
incoherence, frequency locking, and oscillator death, there exist
novel hybrid solutions that combine two or more of these states.
We present the phase diagram and derive several of the stability
boundaries analytically.
\end{abstract}


]

\newpage

The collective behavior of limit-cycle oscillators was first
investigated by Winfree~\cite{winfree67}. Using a mean-field
model of coupled phase oscillators with distributed natural
frequencies, he discovered that collective synchronization is a
threshold phenomenon, the temporal analogue of a phase transition.
Specifically, when the strength of the coupling exceeds a
critical value, some oscillators spontaneously synchronize to a
common frequency, overcoming the disorder in their natural
frequencies.  The model was subsequently refined by
Kuramoto~\cite{kuramoto} and others~\cite{sakaguchi,stroreview},
with applications to Josephson junction arrays~\cite{wiesenfeld},
neutrino flavor oscillations~\cite{pantaleone}, Brownian
ratchets~\cite{haubler}, bubbly fluids~\cite{russo},
semiconductor laser arrays~\cite{kozyreff} and Landau damping of
plasmas~\cite{strogatz92}.

Despite all the activity that Winfree's work ultimately provoked,
surprisingly little is known about the dynamics of his original
model. In this Letter, we explore a special case of the model for
which several analytical results can be obtained. This version of
the model is also related to recent work on pulse-coupled
oscillators (where the oscillators interact by firing sudden
impulses), a case of interest in neurobiology~\cite{bressloff}.
However, we are motivated here by a dynamical systems
perspective: the goal is to understand the collective behavior
and bifurcations of the model as a function of two parameters,
the coupling strength and the spread of natural frequencies.

In the limit of weak coupling and nearly identical frequencies,
our model reduces to the Kuramoto model, whose behavior is well
understood~\cite{kuramoto,sakaguchi,stroreview}: it displays
locked, partially locked, or incoherent states, depending on the
choice of parameters. Away from this familiar regime, we discover
novel hybrid states corresponding to various mixtures of locking,
incoherence, and oscillator death (a cessation of oscillation
caused by excessively strong coupling~\cite{kopell}).

The Winfree model is
\begin{equation}
\dot\theta_i=\omega_i+\frac{\kappa}N\sum_{j=1}^N P(\theta_j)
R(\theta_i), \label{eq1}
\end{equation}
for $i=1,\ldots,N$, where $N\gg 1$. Here $\theta_i(t)$ is the
phase of the $i$th oscillator at time $t$, $\kappa\ge 0$ is the
coupling strength, and the frequencies $\omega_i$ are drawn from a
symmetric, unimodal density $g(\omega)$. We assume that the mean
of $g(\omega)$ equals 1, by a suitable rescaling of time, and
that its width is characterized by a parameter $\gamma$. The
coupling in (\ref{eq1}) has the following interpretation. The
$j$th oscillator makes its presence felt through an influence
function $P(\theta_j)$; in turn, the $i$th oscillator responds to
the mean field (the average influence of the whole population)
according to a sensitivity function $R(\theta_i)$.

From now on, we consider the special case where
\begin{equation}
P(\theta)=1+\cos\theta,\quad R(\theta)=-\sin\theta. \label{eq2}
\end{equation}
Note that this $P(\theta)$ is a smooth but pulse-like function.
(At the end of this paper we consider a much more sharply peaked
$P(\theta)$; then $\theta = 0$ represents the phase when the
oscillator suddenly fires.) The functional form of $R(\theta)$ is
inspired by the qualitative shape of the phase-response curve of
many biological oscillators~\cite{kopell,winfree80}.  With these
choices, Eq.~(\ref{eq1}) becomes a simple model for a population
of pulse-coupled biological oscillators, such as
crickets~\cite{walker}, fireflies~\cite{buck}, or heart pacemaker
cells~\cite{peskin}. The unusual aspect is that the coupling is
through the phase-response curve~\cite{stein}; thus an oscillator
can be either advanced or delayed by a pulse from another
oscillator, depending on its phase when it receives the stimulus.
This differs from the strictly excitatory or inhibitory coupling
often used in integrate-and-fire models of neural oscillators.

We begin by describing our numerical simulations. Equation
(\ref{eq1}) was integrated numerically using $N=800$ oscillators.
The frequencies $\omega_i$ where chosen to be evenly spaced in
the interval $I=[1-\gamma,1+\gamma]$, corresponding to a uniform
density $g(\omega)=1/2\gamma$ for $\omega\in I$, and $g(\omega)=0$
otherwise. The long-term behavior of the system was always found
to be independent of the initial conditions.

To compare the long-term dynamics of individual oscillators, we
compute the average frequency (also known as the rotation number)
of each oscillator, $\rho_i~=~\lim_{t\to\infty}\theta_i(t)/t$. In
our simulations, the limit was typically well-approximated by
integrating the system for 500 time units, starting from a random
initial condition, although longer runs were sometimes necessary.
The rotation numbers provide a convenient measure of
synchronization: two or more oscillators are frequency locked if
they have the same rotation number.

Figure~1 plots $\rho_i$ as a function of $\omega_i$ for
increasing values of $\gamma$, at a fixed $\kappa$. For small
$\gamma$, all the oscillators are locked [Fig.~1(a)]. They can be
visualized as a pack of particles rotating at the same average
rate around the unit circle, where $\theta_i(t)$ denotes the
angular position of oscillator $i$. As $\gamma$ is increased past
a critical threshold, the coupling is no longer sufficient to
keep all the oscillators mutually entrained. Just above
threshold, the system stays partially locked, with the fastest
oscillators peeling away from the pack, but drifting incoherently
relative to one another~[Fig.~1(b)]. With further increases in
$\gamma$, successively more oscillators peel away until
eventually the entire population is incoherent [Fig.~1(c)]. For
sufficiently large~$\gamma$, the system converges to a state of
partial death in which the slowest oscillators stop moving
altogether, while the faster ones remain incoherent [Fig.~1(d)].

\begin{figure}[tbp]
\begin{center}
\psfig{figure=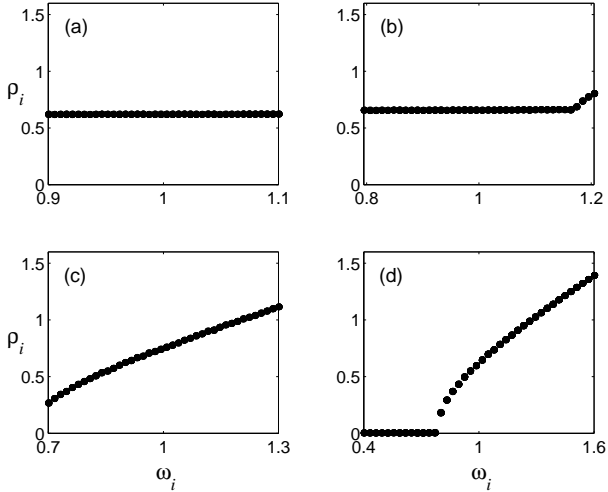,width=3.2in}
\end{center}
\caption{Collective states, as indicated by rotation numbers for
$\kappa=0.65$. Equation~(\ref{eq1}) with $P(\theta)$, $R(\theta)$
as in~(\ref{eq2}) was integrated for 500 time units starting from
a random initial condition. (a)~$\gamma=0.1$: locking.
(b)~$\gamma=0.205$: partial locking. (c)~$\gamma=0.3$:
incoherence. (d)~$\gamma=0.6$: partial death.} \label{fig1}
\end{figure}

\begin{figure}[t]
\begin{center}
\psfig{figure=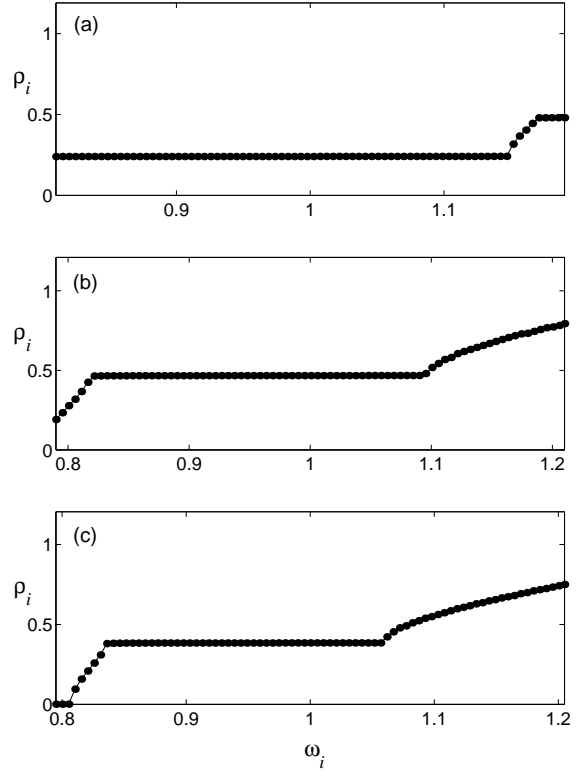,width=3in}
\end{center}
\caption{Partially-locked hybrid states. (a) $\gamma=0.19,$
$\kappa=0.78$: locked-slipping-locked. (b) $\gamma=0.21,$
$\kappa=0.76$: slipping-locked-incoherent.  (c) $\gamma=0.205,$
$\kappa=0.79$: quivering-slipping-locked-incoherent.} \label{fig2}
\end{figure}

Partial locking and partial death are hybrid states; their
rotation number plots [Figs.~1(b) and (d)] contain two distinct
branches that correspond to qualitatively different dynamics. We
have also observed hybrid states with three and four branches
[Fig.~2]. These more complicated states should all be regarded as
variants of partial locking, since there is at least one branch of
locked oscillators in each case. (Note that all these states are
near each other in parameter space.) We label them according to
their branches, as follows. {\em Locked-slipping-locked\/}
[Fig.~2(a)]: There are two separate plateaus of locked
oscillators, at close to 2:1 frequency ratio in the example
shown, separated by a branch of slipping oscillators. A slipping
oscillator typically co-rotates with a locked group for several
periods, then slips away for a few cycles before eventually
rejoining the same group and repeating the pattern. Oscillators
slip more or less frequently depending on their native frequency
$\omega_i$. {\em Slipping-locked-incoherent\/} [Fig.~2(b)]: There
is a central group of locked oscillators, flanked by slower ones
that slip and faster ones that drift monotonically. {\em
Quivering-slipping-locked-incoherent\/} [Fig.~2(c)]: This state
exists near partial death. It is similar to the state in
Fig.~2(b), but with an added mode of behavior: the slowest
oscillators quiver about their former death phases. An oscillator
that quivers has zero rotation number---it remains trapped in the
neighborhood of a single phase for all time---and hence is
effectively dead, though not completely motionless. Figure 3
summarizes the system's long-term behavior as a function of
$\kappa$ and~$\gamma$.

\begin{figure}
\begin{center}
\psfig{figure=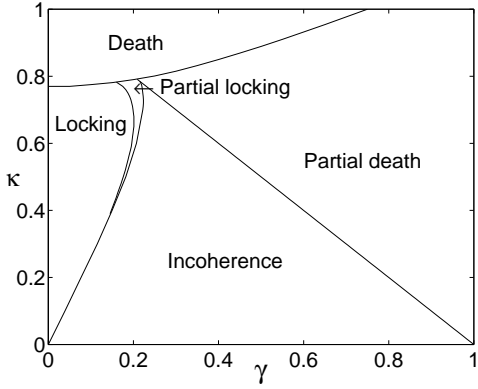,width=2.5in}
\end{center}
\caption{Phase diagram for Eqs.~(\ref{eq1}) and (\ref{eq2}),
assuming a uniform distribution of natural frequencies on
$[1-\gamma,1+\gamma]$. The boundary between locking and partial
locking is determined numerically; all other boundaries are
determined analytically. All of the partially-locked hybrid
states are lumped together in one region, for simplicity.}
\label{fig3}
\end{figure}

We now outline our analytical calculations of the boundaries of
the regions corresponding to incoherence, partial death, and
death.  Following the standard approach used for the Kuramoto
model~\cite{kuramoto,sakaguchi,stroreview}, we rewrite the
dynamics in the infinite-$N$ limit. By analogy with the continuum
limit in fluid mechanics, we view the oscillators as particles
moving around the unit circle. For each frequency $\omega$, let
$p\left(\theta,t,\omega\right)$ denote the density of oscillators
at phase $\theta$ at time $t$, and let $v(\theta,t,\omega)$
denote the local velocity field. Then $p$ satisfies the
continuity equation $\partial p/\partial t = -\partial
\left(pv\right)/\partial\theta$, expressing conservation of
oscillators of frequency $\omega$. The velocity $v$ is determined
by applying the law of large numbers to Eq.~(\ref{eq1}). The sum
over all oscillators in~(\ref{eq1}) is replaced by an integral as
$N\to\infty$, yielding
$v(\theta,t,\omega)=\omega-\sigma(t)\sin\theta$, where
\begin{equation}
\sigma(t)=
\kappa\int_0^{2\pi}\int_{1-\gamma}^{1+\gamma}\left(1+\cos\theta\right)
p\left(\theta,t,\omega\right) g(\omega) d\omega d\theta.
\label{eq3}
\end{equation}
Additionally, we demand that $p$ be non-negative, $2\pi$-periodic
in $\theta$, and we impose the normalization
$\int_0^{2\pi}\!p(\theta,t,\omega)d\theta=1$ for all $t,\omega$.

The key to the analysis is recognizing that incoherence, partial
death, and death correspond to {\em stationary\/} densities
$p_0(\theta,\omega)$ in the infinite-$N$ limit. Hence, one may
solve for all three states by seeking fixed points of the
continuity equation. These satisfy $p_0v_0=C(\omega)$, with
$C(\omega)$ determined by normalization. Depending on its natural
frequency, an oscillator's steady-state behavior falls into one of
two categories. (i) $\omega\le\sigma$: In this case,
$v_0=\omega-\sigma\sin\theta=0$. The oscillators of a given
frequency $\omega\le\sigma$ all remain stuck at a single phase
$\theta^*(\omega)$, defined implicitly by
$\sin\theta^*~=~\omega/\sigma$. Their density is
$p_0(\theta,\omega)=\delta(\theta-\theta^*(\omega))$. Such
oscillators are motionless, or dead. (ii)~$\omega>\sigma$: These
oscillators rotate non-uniformly around the circle, hesitating
near $\theta=\pi/2$ and accelerating near $\theta=3\pi/2$, as
dictated by their velocity field $v_0=\omega-\sigma\sin\theta>0$.
The stationary density is inversely proportional to the velocity:
\begin{equation}
p_0(\theta,\omega)=\frac{C(\omega)}{\omega-\sigma\sin\theta}.
\label{eq4}
\end{equation}
Normalization then implies
$C(\omega)=\sqrt{\omega^2-\sigma^2}/(2\pi)$. From these two basic
scenarios, we are able to calculate the incoherence, partial
death and death boundaries as follows.

Incoherence exists provided $\omega>\sigma$ for all $\omega$; then
all oscillators belong to category (ii) above. The boundary
separating incoherence and partial death occurs when
$\sigma~=~\omega_{min}$, where $\omega_{min}~=~1-\gamma$. The
first oscillators to die are the slowest ones. To solve for
$\sigma$, we substitute (\ref{eq4}) into Eq.~(\ref{eq3}); this
yields $\sigma~=~\kappa$. Thus, partial death bifurcates from
incoherence along the straight line
\begin{equation}
\kappa=1-\gamma, \label{eq5}
\end{equation}
assuming $\kappa$ is not so large that the system is in the death
region. Remarkably, this result holds for any frequency
distribution $g(\omega)$, whether symmetric or not.

The stability of the incoherent state is determined by linearizing
the continuity equation about the incoherent density~(\ref{eq4}).
The resulting linear operator has a continuous spectrum that is
pure imaginary and a discrete spectrum that is governed by the
equation \cite{aria00b}
\begin{equation}
\kappa=\int_{1-\gamma}^{1+\gamma}\frac{\lambda\,\omega\left(\omega-\sqrt{\omega^2-\kappa^2}\right)}
{\lambda^2+\omega^2-\kappa^2}\,g(\omega) d\omega. \label{eq6}
\end{equation}
From (\ref{eq6}), it is clear there are no eigenvalues $\lambda$
with $\mathrm{Re}(\lambda)<0$; if there were, the right-hand side
would have negative real part, contradicting the assumption
$\kappa\ge 0$. Thus, incoherence is either unstable or neutrally
stable. Numerics indicate that the boundary between incoherence
and partial locking corresponds to a Hopf bifurcation. To obtain
the boundary, we solve Eq.~(\ref{eq6}) for $\lambda$
perturbatively, assuming $\gamma\ll 1$, and then take the limit
$\mathrm{Re}(\lambda)\to 0^+$. The result for a uniform
$g(\omega)$ is
\begin{equation}
\kappa=\frac{8\gamma}{\pi}\left[1+\frac{16\gamma^2}{\pi^2}+\frac{16(\pi^2+80)\gamma^4}{\pi^4}\right]+O(\gamma^7).
\label{eq7}
\end{equation}
In the limit $\gamma\to 0$, Eq.~(\ref{eq7}) reduces to
$\kappa=8\gamma/\pi$, which is the critical coupling threshold
for the averaged system (the Kuramoto model, with coupling
$K=\kappa/2$). Hence, Eq.~(\ref{eq7}) can be viewed as an
extension of the classical threshold
condition~\cite{winfree67,kuramoto} into the non-averaged regime
of stronger coupling and frequency disorder.

Finally, to calculate the death boundary we use a self-consistency
argument familiar from mean-field theory~\cite{kuramoto}. In the
death state, each oscillator comes to rest at $\theta^*(\omega)$,
defined by the zero velocity condition
$\sin\theta^*=\omega/\sigma$. This requires $\sigma\ge
\omega_{max}$, where $\omega_{max}=1+\gamma$. Each phase
$\theta^*(\omega)$ depends on $\sigma$, which in turn depends on
all phases via Eq.~(\ref{eq3}); therefore, $\sigma$ must be
determined self-consistently. For each $\omega$, there are two
possible roots $\theta^*(\omega)$: one in $[0,\pi/2]$, the other
in $[\pi/2,\pi]$. However, the unique stable fixed point of
Eq.~(\ref{eq1}) satisfies $0\le\theta^*(\omega)\le\pi/2$ for all
$\omega$. Substituting the corresponding density
$p_0(\theta,\omega)=\delta(\theta-\theta^*(\omega))$ into
Eq.~(\ref{eq3}) gives the self-consistency condition
\begin{equation}\label{eq8}
\frac{\sigma}\kappa=1+\int_{1-\gamma}^{1+\gamma}
\sqrt{1-\left(\frac{\omega}\sigma\right)^2}g(\omega)d\omega.
\end{equation}
For death to exist, there must be a root $\sigma\ge 1+\gamma$ of
Eq.~(\ref{eq8}). The boundary between death and partial death
corresponds to an endpoint bifurcation, and is found by setting
$\sigma=1+\gamma$ in (\ref{eq8}). For a uniform $g(\omega)$, this
yields the exact expression
\begin{equation}
\frac{1}\kappa=\frac{1}{4\gamma}\left[2+\frac{\pi}2-\Delta\left(2+\sqrt{1-\Delta^2}\right)-\sin^{-1}\Delta\right],
\label{eq9}
\end{equation}
where $\Delta=(1-\gamma)/(1+\gamma)$. The remaining portion of the
boundary, separating death from full and partial locking,
corresponds to a saddle-node bifurcation. We obtain it by solving
Eq.~(\ref{eq8}) numerically, together with the tangency condition
$1/\kappa=F'(\sigma)$, where $F(\sigma)$ is the right-hand side
of Eq.~(\ref{eq8}).

To check the robustness of the phase diagram, we replaced
$P(\theta)$ in (\ref{eq2}) with a family of influence functions
$P_n(\theta)~=~a_n(1+\cos\theta)^n$, $n\ge 1$ which becomes more
and more sharply peaked as $n$ increases. (The normalization
coefficients $a_n$ are determined by requiring $P_n(\theta)$ to
have integral equal to $2\pi$ over one cycle. Note that
$P_n(\theta)\to 2\pi\delta(\theta)$ as $n\to\infty$.) We find
that all of the phenomena observed for the model studied in this
paper ($n=1$) persist as we increase $n$; the only difference is
that the boundaries become slightly distorted [Fig.~4].

\begin{figure}
\begin{center}
\psfig{figure=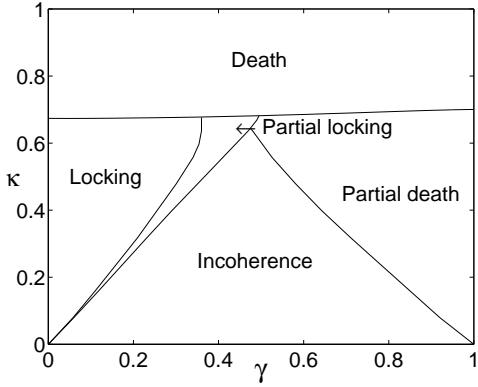,width=2.5in}
\end{center}
\caption{Phase diagram for (\ref{eq1}) with
$P(\theta)=a_n(1+\cos\theta)^n$, $n=10$ and a uniform frequency
distribution on $[1-\gamma,1+\gamma]$. Death boundary determined
analytically; all other boundaries determined numerically.}
\label{fig4}
\end{figure}

The mean-field model (\ref{eq1}) is one of the simplest possible
models of pulse-coupled oscillators. More realistic models would
include such features as spatial coupling, time delay, dynamical
synapses, refractory period, non-sinusoidal influence and
sensitivity functions, etc.  It remains to be seen whether such
models would also exhibit the hybrid states found here.  In any
case, we now know that even the most idealized version of the
Winfree model displays a fascinating wealth of dynamics that,
curiously, escaped notice for over thirty years.

Research supported in part by the National Science Foundation.

\end{document}